\documentclass[12pt]{article}

\usepackage[utf8]{inputenc}
\usepackage{amsmath, amssymb, bm}
\usepackage{graphicx}
\usepackage{booktabs}
\usepackage[flushleft]{threeparttable}
\usepackage{caption}
\usepackage{float}
\usepackage{setspace}
\usepackage{csquotes}
\usepackage[left=1in,right=1in,top=1in,bottom=1in]{geometry}
\usepackage{xurl}   
\usepackage[
  style=apa,
  backend=biber,
  sortcites=true,
  natbib=true
]{biblatex}
\DeclareLanguageMapping{american}{american-apa}
\usepackage[
  colorlinks=true,
  linkcolor=black,
  citecolor=black,
  urlcolor=black
]{hyperref}

\title{Small Area Bayesian Dynamic Borrowing: Adaptive Subgroup Estimation for Large-Scale Educational Assessments}
\author{Sinan Yavuz\thanks{Department of Educational Psychology, University of Wisconsin--Madison. Sinan Yavuz is now a Senior Research Scientist at Amazon; this work was completed before he joined Amazon. Correspondence: \href{mailto:yavuzsinan@gmail.com}{yavuzsinan@gmail.com}.}
\and David Kaplan\thanks{Department of Educational Psychology, University of Wisconsin--Madison.}}
\date{}

\begin{document}

\maketitle
\thispagestyle{empty}

\doublespacing

\begin{abstract}
\noindent
Large-scale assessments suppress subgroup achievement estimates below minimum sample size thresholds, such as the National Assessment of Educational Progress (NAEP) rule of 62, disproportionately affecting historically underrepresented groups. This study introduces Small Area Bayesian Dynamic Borrowing (SABDB), a unit-level small area estimation method assigning each regression coefficient its own between-area variance, so cross-area borrowing adapts to each coefficient's heterogeneity. We compare SABDB against the unit-level Hierarchical Bayesian Small Area Estimation (HBSAE) model in a simulation study and an empirical case study. The simulation mimics the NAEP eighth-grade mathematics assessment, and the case study uses the PISA 2018 dataset. Across both studies, SABDB achieved near-nominal coverage, whereas HBSAE's intervals were narrow but severely miscalibrated.
\end{abstract}


\newpage

\section{Introduction}

Accurately estimating subgroup performance in large-scale educational assessments (LSAs) is a fundamental challenge in educational measurement. The National Assessment of Educational Progress (NAEP), often referred to as the `Nation's Report Card,' provides critical insights into student achievement across states and demographic subgroups. However, a key limitation of NAEP is its reporting threshold policy, which requires a minimum sample size of 62 students (also known as the rule of 62) per subgroup before mean achievement estimates are published \parencite{jones2004nation}. This threshold was established to ensure statistical reliability, aiming to detect an effect size of 0.5 with a probability of 0.8 or greater while accounting for design effects inherent in complex sampling methodologies \parencite{NCES2025}. International assessments apply analogous rules. The OECD requires at least 30 students and 5 schools for subgroup reporting in the Programme for International Student Assessment (PISA), and it cautions against overinterpreting results near that threshold \parencite{PISA2022vol2}.

\begin{figure}[H]
    \centering
    \includegraphics[width=1\textwidth]{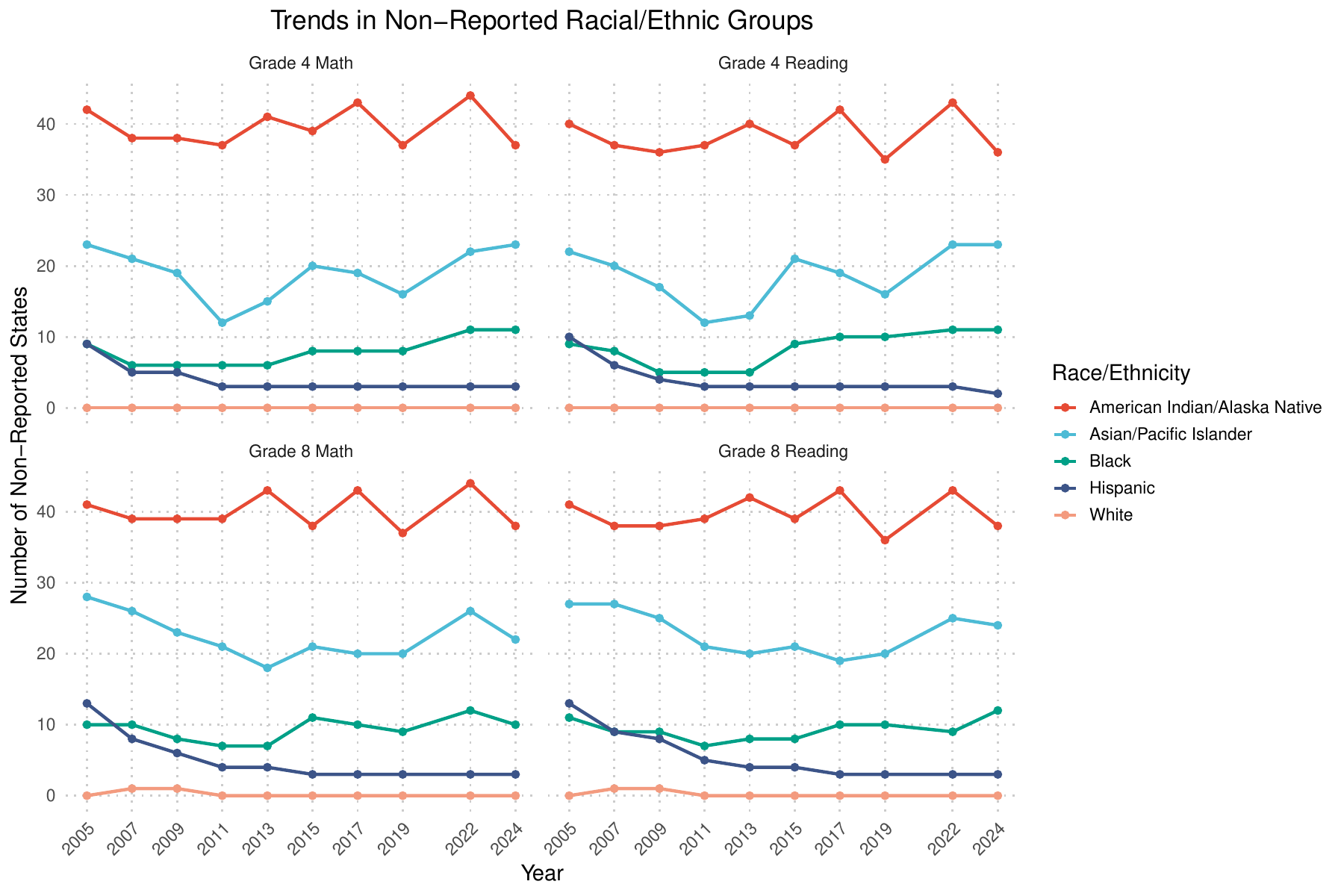}
    \caption{Trends in non-reported racial/ethnic groups in NAEP, Grades 4--8, reading and mathematics, 2005--2024. Each point is the number of states (including the District of Columbia) for which the subgroup mean was not reported. \textit{Source:} U.S. Department of Education, Institute of Education Sciences, National Center for Education Statistics, NAEP 2005--2024 Reading and Mathematics Assessments.}
    \label{fig:non_reported}
\end{figure}

Although intended to protect precision, these policies produce substantial data suppression for exactly the populations most in need of policy attention. As Figure~\ref{fig:non_reported} shows, American Indian/Alaska Native students' achievement has gone unreported in more than 40 states in some assessment years, while Black and Hispanic students face non-reporting in 10--15 states biennially across grade-subject combinations. The suppression obscures the performance of small subgroups in national and state policy discussions, complicating efforts to identify disparities, monitor progress, and allocate resources.

Small area estimation (SAE) responds to this problem by borrowing strength from related domains, so that small samples are stabilized by external information \citep{rao2015small}. Yet traditional SAE methods treat borrowing as a fixed structural assumption, governed by a single variance component shared across all areas. In education, that assumption often fails. Using Stanford Education Data Archive scores linked to state NAEP, \citet{reardon2024separate} document large cross-district heterogeneity in racial achievement gaps. Some districts' White--Black gaps are under 0.2 SD while others approach 1 SD, and gap trajectories from Grades 3 to 8 differ in sign across districts. When subgroup effects vary this much across areas, borrowing uniformly can introduce substantial bias for precisely the marginalized populations whose outcomes diverge from state or national patterns.

This study introduces Small Area Bayesian Dynamic Borrowing (SABDB), a unit-level SAE method that adapts the degree of information borrowing to the data. SABDB extends Bayesian dynamic borrowing (BDB), developed for incorporating historical data in clinical trials \parencite{ibrahim2000power, hobbs2012commensurate, viele2014use} and recently brought to LSAs for borrowing across assessment cycles \parencite{kaplan2023bayesian1, kaplan2023bayesian2}, to borrowing across small area domains. The key feature is that each regression coefficient receives its own between-area variance parameter, so the model borrows strongly for coefficients that are homogeneous across areas and weakly for those that are heterogeneous. The degree of borrowing is governed by the posterior distributions of the variance parameters rather than a pre-specified structure.

We evaluate SABDB in two complementary studies. Study~1 is a simulation using a synthetic population calibrated to NAEP Grade~8 mathematics, where true subgroup means are known by construction. It compares SABDB with a standard hierarchical Bayesian SAE model (HBSAE) under homogeneous and heterogeneous cross-state conditions. Study~2 is an empirical comparison using real PISA 2018 data across 78 countries and 234 country-by-immigration-status domains, where sample sizes range from 2 to over 30,000. It tests whether the borrowing mechanism discovers the heterogeneity structure when it is not known a priori. To our knowledge, this is the first application of Bayesian dynamic borrowing to an international LSA in which countries serve as the areas of a small area estimation framework.

\section{Background}

\subsection{Small Area Estimation}

Small area estimation methods produce reliable estimates for domains (geographic areas, subgroups, or their intersections) whose sample sizes are too small for reliable direct estimation \parencite{rao2015small}. Two model classes dominate practice. Area-level models, beginning with \textcite{fay1979estimates}, relate direct domain estimates to area-level covariates. Unit-level models, beginning with the nested error regression of \textcite{battese1988error}, model individual outcomes with area random effects. Hierarchical Bayesian small-area estimation (HBSAE) implementations assign priors to the hyperparameters and propagate the parameter uncertainty they encode into posterior intervals, generally yielding better-calibrated intervals than empirical Bayes plug-in approaches \parencite{arora1997superiority, ghosh1992hierarchical}. In education, SAE underlies products such as the PIAAC Skills Map of state and county adult skill estimates \parencite{krenzke2020piaac}. Common to these approaches, however, is a fixed borrowing structure: a single between-area variance component governs how much every area, and every aspect of the regression relationship, is pooled toward the center.

\subsection{Bayesian Dynamic Borrowing}

Bayesian Dynamic Borrowing (BDB) is a class of Bayesian statistical methods that incorporate external or historical information into analyzes, adjusting the degree of borrowing based on the compatibility of external data with current data \parencite{viele2014use, kaplan2023bayesian1, kaplan2023bayesian2}. The method borrows heavily when historical data align closely with new data and reduces the influence of external information when substantial differences exist. This adaptation is achieved through hierarchical models or specifically constructed priors that determine the appropriate level of information sharing. Consequently, BDB can improve estimation precision without compromising validity when external data differ from the current context.

The use of BDB in statistical inference has grown with the need to incorporate all available evidence. In fields such as clinical trials, using historical controls or related studies can substantially reduce the required sample size and strengthen conclusions if external information is integrated appropriately \parencite{edwards2024using}. Bayesian methods naturally facilitate such evidence synthesis by treating unknown parameters as random variables with prior distributions that can encode external knowledge. Early applications of borrowing in Bayesian analysis include empirical Bayes shrinkage \parencite{efron1977stein} and Bayesian meta-analysis \parencite{smith1995bayesian}, which demonstrated how pooling information across studies or groups can improve estimation accuracy. However, these earlier approaches often used fixed pooling weights or ad-hoc adjustments. Subsequently, more formal dynamic borrowing techniques were developed. For example, \textcite{pocock1976combination} discussed incorporating historical controls in trials with great caution, and in the decades that followed, statisticians formulated power priors and hierarchical models to include historical data with tunable influence. A key milestone was the introduction of power priors by \textcite{ibrahim2000power}, which allow the likelihood of historical data to be raised to a fractional power $\alpha$ ($0 \leq \alpha \leq 1$) to control its influence. This idea was further refined by \textcite{hobbs2011hierarchical, hobbs2012commensurate}, who proposed commensurate priors, treating the weight for historical data as an unknown parameter to be estimated from the data. Commensurate prior models introduced a hyperparameter to capture the commensurability between current and historical results, making the borrowing dynamic. Building on these advances, \textcite{viele2014use} provided a comprehensive comparison of historical borrowing methods, distinguishing static from dynamic approaches and reviewing methods that incorporate a heterogeneity parameter into the prior linking current and historical data, so that the posterior automatically adjusts the degree of borrowing according to observed differences.

\textcite{kaplan2023bayesian1} extended BDB to the analysis of large-scale educational assessments using single-level and multilevel regression models with covariates, applying it to multiple cycles of PISA data. They showed that BDB performs at least as well as complete pooling when historical data are homogeneous, and better than pooling and power priors when historical data are heterogeneous, in terms of bias, mean squared error, and predictive accuracy. \textcite{kaplan2023bayesian2} further extended BDB to longitudinal large-scale assessments, demonstrating its utility for borrowing across assessment cycles over time.

This article builds on these two studies by extending BDB to the domain of small area estimation. We develop the Small Area Bayesian Dynamic Borrowing (SABDB) method, which applies the dynamic borrowing mechanism across small area domains rather than across historical time points, allowing the degree of information sharing between domains to be determined by the data. BDB methods have been adopted across various fields, from biomedical research \parencite{viele2014use} to educational assessment \parencite{kaplan2023bayesian1}, whenever researchers seek to combine information from past studies with new data. The increasing use of BDB is also reflected in regulatory guidance, with evolving discussions on the acceptability of dynamically borrowing external controls in clinical trials, especially in settings like rare diseases and pediatric studies \parencite{edwards2024using}.

A formal mathematical treatment of BDB, including the hierarchical formulation, its power-prior and effective-sample-size representations, and its limiting behavior as the current sample grows, is provided in the online supplement (Section~S12). The following section develops how BDB is incorporated into small area estimation.

\section{The SABDB Model}\label{sec:sabdb_model}

\subsection{Motivation and Formulation}

Consider estimating mean achievement for a small subgroup domain, such as Black students with an Individualized Education Program in a particular state, from unit-level assessment data, where the subgroup sample is often too small for stable direct estimation. Pooling all states assumes homogeneity that may not hold. A conventional HB model with a single between-state variance may still over-borrow from dissimilar states if the prior structure is too rigid. SABDB instead allows each regression coefficient its own between-area variance, so the degree of borrowing adapts separately for each aspect of the regression relationship.

Let $Y_{s,j}$ denote the achievement score of student $j$ in area $s$ ($s = 1, \dots, S$), and $\mathbf{x}_{s,j}$ a vector of $K$ covariates including an intercept. The area-specific coefficient vector is $\boldsymbol{\beta}_s = (\beta_{s,1}, \dots, \beta_{s,K})^\top$. The model is
\begin{align}
    Y_{s,j} \mid \boldsymbol{\beta}_s, \sigma_y^2 &\sim \mathcal{N}(\mathbf{x}_{s,j}^{\top} \boldsymbol{\beta}_s,\ \sigma_y^2), \label{eq:sabdb_likelihood}\\
    \beta_{s,k} \mid \mu_k, \tau^2_k &\sim \mathcal{N}(\mu_k,\ \tau^2_k), \qquad k = 1, \dots, K, \label{eq:sabdb_prior}
\end{align}
where $\sigma_y^2$ is the within-area residual variance, $\mu_k$ is the global mean of coefficient $k$, and $\tau^2_k$ is the between-area variance for coefficient $k$. The key parameters for dynamic borrowing are the between-area variance parameters $\tau^2_k$. When areas have similar values of coefficient $k$, the posterior for $\tau^2_k$ will concentrate on small values, resulting in strong borrowing for that coefficient. When areas differ substantially, the posterior for $\tau^2_k$ will shift toward larger values, reducing borrowing. The hierarchical prior in Equation~\ref{eq:sabdb_prior} treats the $\beta_{s,k}$ as independent across $k$, so the between-area covariance matrix is diagonal with elements $\tau^2_k$. This follows the specification in \textcite{kaplan2023bayesian1}, where the covariance matrix governing borrowing is diagonal and each element is given its own prior distribution.

All covariates and the outcome are standardized prior to estimation, and posterior summaries are transformed back to the reporting scale. On the standardized scale the hyperpriors are
\begin{align}
    \mu_k &\sim \mathcal{N}(0, 3^2), \label{eq:mu_prior}\\
    \tau^2_k &\sim \text{Inv-Gamma}(1,\ 0.001), \label{eq:tau_prior}\\
    \sigma_y &\sim \text{Half-Cauchy}(0, 5). \label{eq:sigma_prior}
\end{align}
The Inverse-Gamma$(1, 0.001)$ hyperprior is proper, heavy tailed, and concentrated near zero with sufficient mass on larger values to accommodate substantial heterogeneity when the data support it. It was among the specifications evaluated by \textcite{kaplan2023bayesian1}, and its sensitivity is assessed in Study~2 across five alternative specifications. Equation~\ref{eq:sabdb_likelihood} likewise assumes a common residual variance. Area-specific residual variances would add $S$ parameters and were not pursued here.

\subsection{Comparison Model: Hierarchical Bayesian SAE}

The comparison model is the standard unit-level HB SAE model based on the nested error regression framework \parencite{battese1988error}, the most widely used unit-level SAE model in practice \parencite{rao2015small}:
\begin{equation}\label{eq:hbsae}
    Y_{s,j} \mid \boldsymbol{\beta}, u_s, \sigma^2_y \sim \mathcal{N}(\mathbf{x}_{s,j}^{\top} \boldsymbol{\beta} + u_s,\ \sigma^2_y), \qquad u_s \sim \mathcal{N}(0, \sigma^2_u),
\end{equation}
with a single coefficient vector $\boldsymbol{\beta}$ shared across areas and an area random intercept $u_s$. The distinction between SABDB and HBSAE is structural: SABDB estimates $S \times K$ area-specific coefficients governed by $K$ between-area variances, whereas HBSAE estimates $K$ fixed coefficients plus $S$ random intercepts governed by one variance $\sigma^2_u$. HBSAE therefore forces all between-area differences in slopes into the residual term, while SABDB lets the data allocate heterogeneity coefficient by coefficient.

\section{Study 1: Simulation Calibrated to NAEP}

\subsection{Synthetic Population}

This study uses a synthetic population that mimics the National Assessment of Educational Progress (NAEP) student-level data. NAEP student-level data are restricted-use and were not available for this study. A synthetic population was therefore constructed from publicly available NCES data products and calibrated to published NAEP summary statistics. With synthetic data, population parameters are known by construction, so bias, error, and coverage can be evaluated against truth, which is impossible with real data. (Study~2 complements this design with real data.)

The population was built on the Common Core of Data 2022--2023 universe files for state, district, school, and Grade-8 enrollment structure \parencite{NCES_CCD_Data_Files}, with school locale from the NCES EDGE files \parencite{NCES_School_Locations}. State-level targets for subgroup means and proportions came from the 2024 NAEP Grade~8 mathematics Data Explorer and associated technical documentation \parencite{NCES_NDE, NCES_MathConstants_2022}. The data generating model includes the major NAEP reporting variables: race/ethnicity, gender, English Language Learner (ELL) status, Individualized Education Program (IEP) status, National School Lunch Program (NSLP) eligibility, parental education, and school locale. Scores follow the NAEP 0--500 metric with a target population SD of 40 \parencite{NCES_ScoresAchv}, state and school random intercepts yielding intraclass correlations of 0.03 and 0.15 \parencite{yang2024relationship, snijders2012multilevel}, and fixed effects calibrated to reproduce published subgroup differences \parencite{yavuz2024absenteeismncme}. Post-generation alignment steps matched state means, subgroup means, and the population SD to NAEP targets. Quality-control diagnostics show that state-level marginals for ELL, NSLP, and parental education align well with NAEP targets and that race-by-state means track NAEP benchmarks with minimal systematic bias. Full data-generation details, imputation rules for sparse cells, and diagnostic figures appear in the online supplement (Sections~S1--S2).

One property of the data generating model matters for interpretation: it uses fixed regression slopes shared across all states with only random intercepts, a structure that matches HBSAE's assumptions rather than SABDB's. To the extent this alignment matters, the simulation is a conservative test of SABDB.

\subsection{Heterogeneity Scoring and Simulation Conditions}

Because SABDB borrows across states, the design requires a principled way to quantify each state's similarity to the broader pool. For each state we computed a composite between-state heterogeneity score summing three standardized components: distance between the state's race-specific means and national means, distance between the state's racial achievement gaps and national gaps, and divergence between the state's racial composition and the national composition (supplement, Section~S3). After an eligibility screen requiring at least 62 students in each targeted state-subgroup cell, the 10 lowest-scoring states formed the \emph{homogeneous} condition (e.g., Illinois, Arizona, North Carolina) and the 10 highest-scoring states the \emph{heterogeneous} condition (e.g., Alaska, New Mexico, Wisconsin). The heterogeneous set spans distinct dimensions of dissimilarity: compositional divergence (Alaska), unusual gap patterns (Wisconsin), and atypical subgroup means (Oklahoma).

\subsection{Design and Estimation}

Within each condition, 100 independent replications were drawn using a stratified two-stage design: up to 30 schools per state, then up to 30 students per school. Target domains are state-by-race subgroups whose sample sizes fall below the NAEP reporting threshold of 62. Both models were fit in Stan \parencite{carpenter2017stan} using Hamiltonian Monte Carlo with the No-U-Turn Sampler: 4 chains, 4,000 iterations per chain (2,000 warmup), thinned by 10. Across all 400 fits (2 models $\times$ 2 conditions $\times$ 100 replications), no fit produced $\hat{R} > 1.05$ (maximum 1.020). Minimum effective sample size was 387. Only 3 fits, all HBSAE under heterogeneity, produced any divergent transitions.\footnote{Divergent transitions occur when Hamiltonian Monte Carlo's numerical integrator (the leapfrog step) fails to accurately track the true trajectory through the posterior. Intuitively, they flag places where the posterior geometry is too sharp for the sampler to explore reliably, so the affected draws cannot be trusted and often indicate a poorly conditioned posterior.} Design weights were computed but the models were fit unweighted under a model-based inference approach, isolating the borrowing mechanisms from design-based adjustments. This assumes the design is non-informative conditional on model covariates, and pseudo-likelihood extensions are noted in the Discussion \parencite{you2003pseudo, parker2023comprehensive}.

Performance is evaluated against the true population domain means using 95\% coverage probability (CP), mean absolute difference (MAD), root mean squared error (RMSE), and the standard deviation of absolute differences across replications.

\subsection{Results}

\begin{figure}[H]
    \centering
    \includegraphics[width=1\textwidth]{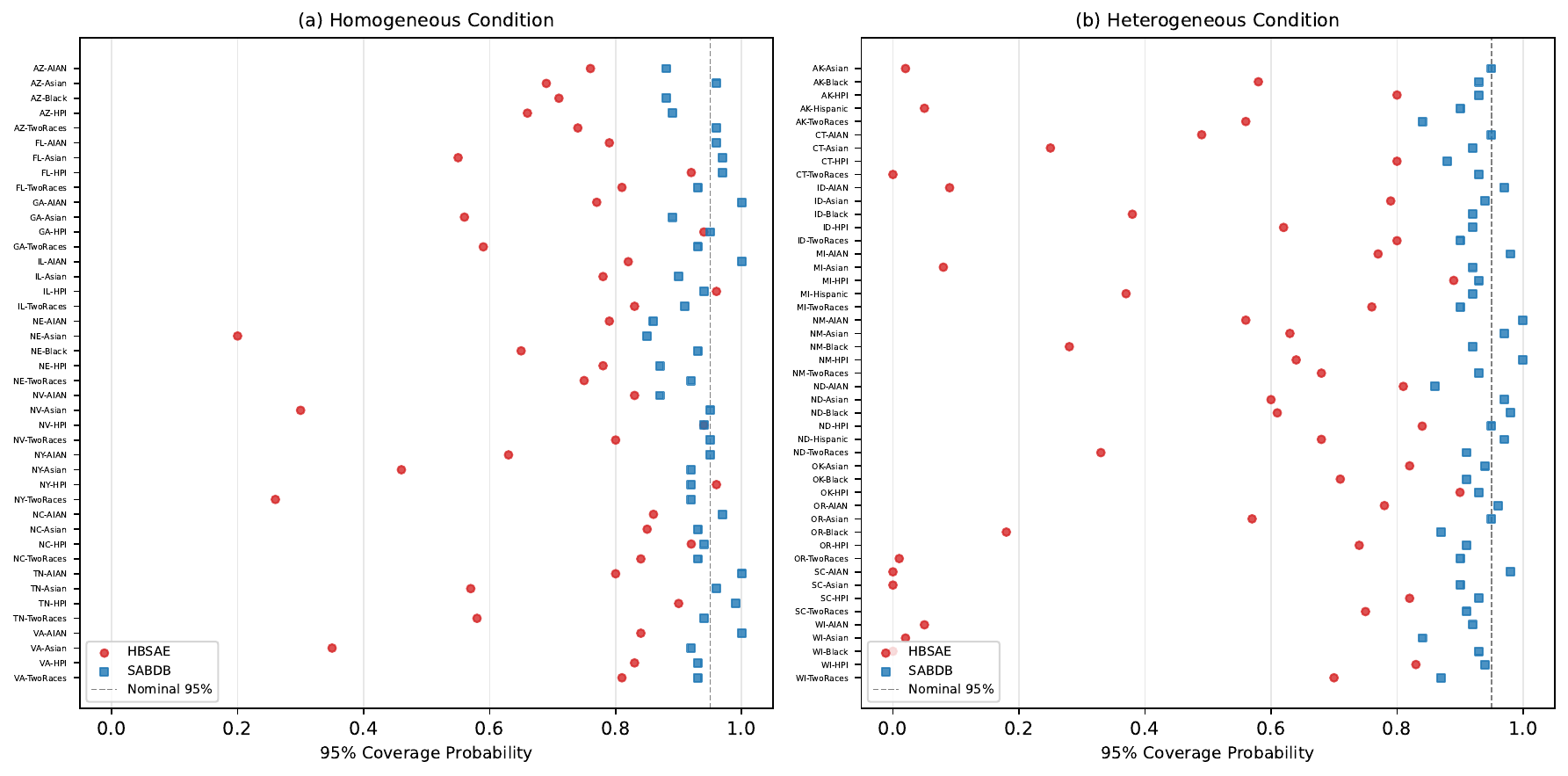}
    \caption{95\% coverage probability for each state-by-subgroup domain under homogeneous and heterogeneous simulation conditions. The dashed line indicates the nominal 95\% level.}
    \label{fig:sim_coverage}
\end{figure}

Table~\ref{tab:sim_summary} summarizes performance, and Figure~\ref{fig:sim_coverage} displays domain-level coverage. SABDB sustains mean 95\% coverage of 0.93 in both conditions, while HBSAE drops from 0.72 under homogeneity to 0.50 under heterogeneity. Per-domain results for all state-by-subgroup cells appear in the supplement (Tables~S4--S5).

\begin{table}[ht]
\centering
\caption{Summary of Model Performance Across Simulation Conditions}
\label{tab:sim_summary}
\begin{tabular}{llcccc}
\toprule
Condition & Model & Mean 95\% CP & MAD & RMSE & SD(MAD) \\
\midrule
Homogeneous & HBSAE & 0.72 & 5.18 & 6.28 & 5.16 \\
Homogeneous & SABDB & 0.93 & 10.80 & 13.63 & 13.38 \\
Heterogeneous & HBSAE & 0.50 & 7.45 & 8.36 & 4.44 \\
Heterogeneous & SABDB & 0.93 & 10.69 & 13.52 & 13.41 \\
\bottomrule
\end{tabular}
\end{table}

\subsubsection{Results under Homogeneous Condition}

A summary of estimator performance for all state--subgroup domains under the homogeneous simulation condition is provided in the online supplement (Table~S4). Across states and racial or ethnic subgroups, the trade-off between precision and coverage under the homogeneous simulation condition was clear. SABDB achieved more consistent nominal 95\% coverage compared to HBSAE, but this improvement came at the cost of higher absolute bias and root mean squared error (RMSE). For instance, among Asian students in Arizona (average sample size approximately 25), coverage increased from 0.69 for HBSAE to 0.96 for SABDB, while the mean absolute bias rose from 3.63 to 6.64 and the RMSE from 4.30 to 8.22. Similar patterns were observed for Asian students in Georgia (coverage 0.56 to 0.89, mean absolute bias 4.43 to 5.48, RMSE 5.35 to 7.30) and for students identifying as Two Races in North Carolina (coverage 0.84 to 0.93, mean absolute bias 2.47 to 4.24, RMSE 3.12 to 5.53). Overall, SABDB produced intervals that captured the true subgroup means more frequently, but with wider uncertainty bands and somewhat less efficient point estimates. This pattern showed that dynamic borrowing approaches enhance interval calibration when subgroup similarity assumptions are satisfied, though often at the expense of slightly increased variance.

The smallest-sample subgroups, particularly Native Hawaiian or Pacific Islander (HPI) populations with average sample sizes of about two students per state, amplified this difference. In Florida, coverage for HPI increased from 0.92 under HBSAE to 0.97 under SABDB, but the mean absolute bias rose from 6.81 to 22.27 and the RMSE from 8.44 to 27.62. Nebraska exhibited the same tendency, with coverage improving from 0.78 to 0.87 while bias and RMSE increased from 9.90 to 30.19 and from 11.80 to 37.49, respectively. In Virginia, HPI coverage improved from 0.83 to 0.93, while bias and RMSE increased from 8.10 to 25.63 and from 10.32 to 31.87. These extremely small-area cases illustrate how heavy regularization improves interval coverage but inflates estimation error when the borrowing source diverges from the target subgroup. This outcome reflects a well-documented phenomenon in small area estimation applications, where hierarchical pooling introduces regression-to-the-mean effects, especially for domains with very limited data \parencite{rao2015small}.

For moderate sample sizes (between approximately 20 and 60 students per subgroup), the coverage advantage of SABDB remained consistent. Examples include Two Races in Arizona (0.74 to 0.96), Florida (0.81 to 0.93), Illinois (0.83 to 0.91), North Carolina (0.84 to 0.93), Nevada (0.80 to 0.95), New York (0.26 to 0.92), Tennessee (0.58 to 0.94), and Virginia (0.81 to 0.93). However, the mean absolute bias and RMSE generally increased as well, such as for Two Races in Illinois (bias 2.54 to 6.65, RMSE 3.30 to 8.13) and Asian students in North Carolina (bias 2.57 to 7.50, RMSE 3.20 to 9.06). This pattern, characterized by higher coverage but higher error, reflects SABDB's conservative borrowing design. The model dynamically adjusts the strength of information sharing and often increases interval width to maintain robustness when subgroup homogeneity is uncertain \parencite{viele2014use, kaplan2023bayesian1, kaplan2023bayesian2}.

Two additional patterns are notable. First, cells with fewer than 100 eligible replications, such as Asian students in New York (13 eligible replications), reflect that fewer replications produced subgroup samples above the reporting threshold of 62, but these cells still exhibit the same general coverage-error trade-off. Second, HBSAE's systematic tendency to pull subgroup means toward the grand mean of each race is evident across states. This shrinkage toward a common mean is a defining characteristic of hierarchical Bayesian estimation, where subgroup estimates are drawn toward the pooled mean in proportion to the ratio of within- to between-group variance. Under homogeneous conditions, this behavior efficiently reduces bias and RMSE but can lead to under-coverage when intervals are too narrow to reflect true sampling variability \parencite{rao2015small, krenzke2020piaac}.

In summary, under homogeneous conditions, SABDB consistently achieved higher 95\% coverage across nearly all state-subgroup combinations, especially among the smallest subgroups. HBSAE produced smaller bias and RMSE but exhibited substantial under-coverage. These findings align with theoretical expectations: HBSAE, whose hierarchical structure parallels the linear data-generating model, provides efficient but overconfident estimates, whereas SABDB trades some precision for improved interval calibration. This trade-off illustrates the principal strength of SABDB: its ability to provide more realistic uncertainty estimates when traditional model assumptions are relaxed. The following subsection examines results under heterogeneous conditions, where we expect SABDB to perform even better because it can adjust borrowing strength and reduce the shrinkage bias that HBSAE exhibited above.

\subsubsection{Results under Heterogeneous Condition}

A summary of estimator performance for all state--subgroup domains under the heterogeneous simulation condition is provided in the online supplement (Table~S5). The results under the heterogeneous condition revealed substantial differences in estimator behavior compared to the homogeneous setting. Across nearly all state-by-subgroup domains, SABDB achieved markedly higher 95\% coverage than HBSAE. In many cases, HBSAE exhibited extremely low coverage, sometimes approaching zero, while SABDB maintained values near the nominal level. For example, in the Alaska--Asian domain, where the average sample size is approximately 26, HBSAE coverage was 0.02, whereas SABDB coverage reached 0.95. Similar improvements were observed in Connecticut--Asian (0.25 to 0.92), Oregon--Two Races (0.01 to 0.90), and South Carolina--AIAN (0.00 to 0.98). These outcomes indicated that SABDB responded well to cross-state heterogeneity by limiting borrowing from dissimilar domains and producing credible intervals that reflect the increased variability in the data. This advantage was clearest when subgroup structures differ substantially across states.

The heterogeneous condition also highlighted important limitations of the traditional HBSAE model. In many domains, HBSAE's global pooling structure did not adjust sufficiently to state-specific differences, resulting in excessive shrinkage towards an overall subgroup mean. When subgroup effects differ across states, this shrinkage could lead to substantial estimation errors. This pattern was evident in domains such as Alaska--Hispanic, Connecticut--Two Races, and Wisconsin--Black, where HBSAE coverage collapsed while mean absolute bias and RMSE remained high. SABDB, in contrast, often exhibited lower bias than HBSAE in these cases. For instance, in Alaska--Hispanic, mean absolute bias decreased from 11.20 under HBSAE to 5.49 under SABDB, and in Michigan--Asian, bias decreased from 12.32 to 7.81. These results showed that SABDB better differentiated between similar and dissimilar borrowing sources, reducing the influence of inappropriate pooling and improving estimation accuracy in heterogeneous settings.

Performance differences between the two models were also notable in domains with extremely small sample sizes. Under these conditions, both HBSAE and SABDB faced estimation challenges, yet SABDB generally provided better-calibrated intervals. For example, in South Carolina--HPI, a domain with an average sample size of approximately two, SABDB improved coverage from 0.82 to 0.93. Although this improvement came with an increase in mean absolute bias, the behavior was consistent with SABDB's conservative borrowing strategy, which favors interval calibration over strict precision when subgroup heterogeneity is high. This trade-off appeared across several low-sample domains, including Idaho--HPI and Alaska--HPI, consistent with SABDB's design, which widens intervals where borrowing strength is uncertain. This conservative treatment enhanced coverage and stability in settings prone to extreme sampling variability.

In subgroups with moderate sample sizes, typically between 15 and 40 students, SABDB often improved both interval calibration and point accuracy. For example, in Alaska--Asian, SABDB reduced mean absolute bias from 18.58 to 6.78 and RMSE from 19.16 to 8.34. In Michigan--Hispanic, coverage increased from 0.37 to 0.92 with only a modest increase in bias, representing a more balanced performance profile. Similarly, in Oregon--TwoRaces, the mean absolute bias decreased substantially from 11.50 to 4.11 under SABDB. These cases demonstrated the strength of the dynamic framework. When subgroup means diverge moderately but not severely, SABDB used the available similarity information to borrow selectively, producing more efficient estimates without compromising coverage.

The results indicated that SABDB outperformed HBSAE across heterogeneous data-generating conditions. HBSAE's fixed hierarchical structure, which presumes a shared linear mean structure across states, limited its flexibility in the presence of subgroup heterogeneity. This often led to severe under-coverage and biased estimates. SABDB, by contrast, adjusted borrowing strength based on empirical evidence of similarity, producing credible intervals that remained stable and well calibrated even as subgroup structures deviate across states. The greatest gains for SABDB appeared in states with strong subgroup-level divergence, such as Alaska, Oregon, and Wisconsin, showing that SABDB could adapt to complex population structures.

\subsection{Decomposition of Error by Sample Size}

The aggregate MAD and RMSE values in Table~\ref{tab:sim_summary} mask sharply different behavior across sample-size strata. Table~\ref{tab:band_decomposition} decomposes error by average per-cell sample size.

\begin{table}[ht]
\centering
\caption{MAD and RMSE Decomposition by Average Cell Sample Size, Study 1}
\label{tab:band_decomposition}
\small
\begin{tabular}{lrrrrr}
\toprule
& & \multicolumn{2}{c}{MAD} & \multicolumn{2}{c}{RMSE} \\
\cmidrule(lr){3-4} \cmidrule(lr){5-6}
Sample size band & Cells & HBSAE & SABDB & HBSAE & SABDB \\
\midrule
\multicolumn{6}{l}{\textit{Homogeneous condition}} \\
\quad $n \le 3$              & 14 & 6.81 & 19.28 & 8.36 & 24.46 \\
\quad $3 < n \le 10$         &  3 & 5.80 & 10.04 & 7.03 & 12.71 \\
\quad $10 < n \le 30$        & 11 & 4.52 &  7.49 & 5.37 &  9.36 \\
\quad $n > 30$               & 14 & 3.94 &  5.08 & 4.76 &  6.35 \\
\quad \textit{Aggregate}     & 42 & 5.18 & 10.80 & 6.28 & 13.63 \\
\midrule
\multicolumn{6}{l}{\textit{Heterogeneous condition}} \\
\quad $n \le 3$              & 10 & 8.49 & 22.26 & 9.90 & 28.08 \\
\quad $3 < n \le 10$         &  8 & 5.55 & 11.27 & 6.50 & 14.63 \\
\quad $10 < n \le 30$        & 15 & 8.74 & \textbf{7.78} & 9.53 & 9.87 \\
\quad $n > 30$               & 14 & 6.42 & \textbf{5.21} & 7.08 & \textbf{6.40} \\
\quad \textit{Aggregate}     & 47 & 7.45 & 10.69 & 8.36 & 13.52 \\
\bottomrule
\end{tabular}
\end{table}

Two patterns are notable. First, the aggregates are dominated by near-empty cells ($n \le 3$, all Native Hawaiian/Pacific Islander and American Indian/Alaska Native subgroups), where neither method receives appreciable within-cell information and SABDB's similarity-weighted borrowing target is itself estimated from sparse data. In these cells HBSAE's heavy shrinkage toward the global subgroup mean produces estimates close to the truth largely by construction. The data generating model placed true means in these cells near the global mean, so global shrinkage performs well. The same outcome would not be expected where true subgroup means deviate from the global mean. Second, among domains with more than 10 sampled students, all still below the reporting threshold of 62, SABDB outperforms HBSAE on accuracy under heterogeneity. For $n > 30$, SABDB attains lower MAD (5.21 vs.\ 6.42) and lower RMSE (6.40 vs.\ 7.08), and for $10 < n \le 30$ lower MAD (7.78 vs.\ 8.74) with comparable RMSE. A complementary analysis found that SABDB's empirical standard error scales with $n$ at approximately the same rate as the direct standard error, so the calibration gains here do not come with a precision gain over direct estimation. The conditions under which SABDB does narrow intervals relative to direct estimation are documented in Study~2.

However, simulation studies are inherently limited by the assumptions embedded in the data-generating process. When the DGM shares structural features with the analytic model being evaluated, favorable results may reflect alignment between the two specifications rather than genuine robustness. This concern motivates the real-data empirical comparison presented in the following section, where SABDB is applied to PISA 2018 international assessment data across 78 countries. The PISA application tests whether the advantages observed in the simulation translate to a setting where the true data-generating process is unknown and potentially far more complex than any parametric model.

\section{Study 2: Empirical Comparison with PISA 2018}

\subsection{Rationale and Setting}

Real data exhibit complexities that a synthetic data-generating mechanism cannot fully reproduce: higher-order dependencies, nonlinearity, heteroscedasticity across clusters, and patterns of missingness. Moreover, the key innovation of SABDB, coefficient-specific between-area variances, was given its heterogeneity structure in the simulation. With real data, the pattern of heterogeneity is unknown and must be discovered by the model.

\subsection{The Programme for International Student Assessment}

The Programme for International Student Assessment (PISA) is a triennial international survey coordinated by the Organisation for Economic Co-operation and Development \parencite{PISA2018}. The program assesses the skills and knowledge of 15-year-old students in mathematics, reading, and science, with the goal of evaluating education systems worldwide. Since its first administration in 2000, PISA has become the largest international comparative assessment of student achievement. The 2018 cycle included approximately 600,000 students from 79 countries and economies, representing over 32 million 15-year-olds \parencite{pisa2018techreport}.

The Programme for International Student Assessment differs from national assessments such as NAEP in several important respects. First, PISA assesses `literacy' rather than curriculum-based knowledge, focusing on students' ability to apply what they have learned to real-world problems \parencite{PISA2018}. Second, while both PISA and NAEP use two-stage stratified sampling designs and rotated test booklets with plausible values, the two programs differ in scope and purpose. PISA samples 15-year-olds across entire national education systems, whereas NAEP samples students within specific grades in the United States. Similar to NAEP, in PISA, schools are sampled with probability proportional to size in the first stage, and a target number of 15-year-old students are randomly selected within each sampled school in the second stage \parencite{pisa2018techreport}. Like NAEP, PISA also uses a latent regression model that produces multiple plausible values rather than individual point estimates \parencite{mislevy1992estimating, wu2005plausible}. The 2018 cycle provides 10 plausible values for each cognitive domain.

The PISA sampling design produces nationally representative samples within each country, but the sample sizes for specific subpopulations can vary enormously. This is particularly true for immigrant subgroups, as native students typically number in the thousands per country while first-generation and second-generation immigrant students may number in the single digits in countries with low immigration rates. This feature makes PISA a natural testing ground for small area estimation methods.

\subsection{Immigration and Educational Achievement in PISA}

Immigration and its relationship to educational outcomes has been a central theme in PISA research since the program's inception. The OECD has devoted substantial attention to this topic, including a dedicated report on the resilience of students with immigrant backgrounds \parencite{OECD2018resilience} and a full volume of the PISA 2018 results examining equity and inclusion \parencite{PISA2018vol2}. These analyses have documented several consistent findings across PISA cycles.

First, the achievement gap between immigrant and native students varies enormously across countries, reflecting differences in immigration policies, the socioeconomic composition of immigrant populations, the quality of integration programs, and the structure of educational systems. In traditional immigration countries such as Canada and Australia, immigrant students perform at or above the level of native students, while in many European countries, substantial gaps persist even after controlling for socioeconomic status.

Second, socioeconomic status is a powerful mediator of the immigration-achievement relationship. The PISA Index of Economic, Social and Cultural Status (ESCS) explains a substantial portion of the achievement gap in most countries, but the degree of mediation varies across national contexts. This variation implies that the relationship between immigration status, socioeconomic status, and achievement is not well characterized by a single set of regression coefficients applied uniformly across all countries.

Third, the small sample sizes of immigrant subgroups in many countries limit the precision of country-specific estimates. Standard PISA reporting presents results only for subgroups that meet minimum sample size thresholds, leaving many country-by-immigration-status combinations unreported. This is precisely the problem that small area estimation methods are designed to address, borrowing strength across countries to improve estimation for domains where direct estimates are unreliable.

\subsection{Data and Variables}

\subsubsection{Data Source and Filtering}

The analysis uses the PISA 2018 student-level data, accessed through the \texttt{EdSurvey} R package \parencite{edsurvey}. After filtering to students with valid immigration status classifications, the initial sample comprises 574,187 students nested within 78 countries. After further excluding students with missing values on the ESCS covariate, the final analysis sample is 571,251 students.

Students are classified into three immigration subgroups based on the PISA-derived variable \texttt{IMMIG}: (1) \textit{Native} students, defined as those with at least one parent born in the country of assessment; (2) \textit{Second-generation} students, born in the country of assessment but whose parents were both born in another country; and (3) \textit{First-generation} students, born outside the country of assessment with parents also born in another country. These three subgroups crossed with 78 countries yield 234 country-by-immigration-status domains, which constitute the small areas of interest.

Table~\ref{tab:pisa_sample} presents the overall sample distribution by immigration status. Native students constitute the vast majority of the sample (87.5\%), while first-generation and second-generation students each represent approximately 6\% of the total. The weighted mean mathematics scores differ across subgroups, with second-generation students scoring highest on average (476.4) and native students scoring lowest (442.2), though these aggregate figures mask substantial heterogeneity across countries.

\begin{table}[ht]
\centering
\caption{PISA 2018 Sample Distribution by Immigration Status}
\label{tab:pisa_sample}
\begin{tabular}{lrrcc}
\toprule
\textbf{Subgroup} & \textbf{N} & \textbf{Percent} & \textbf{Weighted Mean Math} & \textbf{Weighted SD} \\
\midrule
Native & 502,542 & 87.5\% & 442.2 & 106.1 \\
First-Generation & 35,566 & 6.2\% & 453.5 & 106.1 \\
Second-Generation & 36,079 & 6.3\% & 476.4 & 97.7 \\
\midrule
\textbf{Total} & \textbf{574,187}\footnotemark & & & \\
\bottomrule
\end{tabular}
\end{table}
\footnotetext{After excluding students with missing immigration status and students from Japan. The final analysis sample after removing missing ESCS values is 571,251.}

Weighted mean mathematics scores by immigration status for all 78 countries, sorted by native student achievement, appear in the online supplement (Section~S6). They reveal substantial cross-country variation in both overall achievement levels and the magnitude and direction of immigration-related gaps.

\subsubsection{Variables}

The outcome variable is the PISA 2018 mathematics assessment, measured on the PISA scale with an OECD mean of approximately 500 and a standard deviation of approximately 100. PISA reports proficiency as 10 plausible values (PVs) per student, reflecting the latent nature of the construct. To properly account for measurement uncertainty, both SABDB and HBSAE were fit separately to each of the 10 plausible values, and the posterior draws were pooled across all 10 fits. This is the Bayesian analog of Rubin's combining rules \parencite{rubin1987multiple}: instead of combining point estimates and variances algebraically, one concatenates the MCMC draws from each PV-specific fit. The combined posterior automatically captures both model uncertainty (within each PV fit) and measurement uncertainty (across PV fits). This approach follows the recommendation of \textcite{ZhouReiter} for Bayesian inference with multiply imputed data and was also noted as a direction for future work in \textcite{kaplan2023bayesian1}. All domain-level results reported below use the combined 10-PV posteriors. For the direct estimates, we apply Rubin's rules \parencite{rubin1987multiple} to combine information across all 10 plausible values. For each domain, the combined point estimate $\bar{Q}$ is the mean of the 10 plausible-value-specific weighted means, and the total variance is:
\begin{equation}
T = \bar{U} + \left(1 + \frac{1}{M}\right) B
\end{equation}
where $\bar{U}$ is the mean within-PV sampling variance, $B$ is the between-PV variance, and $M = 10$ is the number of plausible values.

The covariate included in the model is the PISA Index of Economic, Social and Cultural Status (ESCS), a continuous composite measure derived from parental occupation (HISEI), parental education (PARED), and home possessions (HOMEPOS) \parencite{pisa2018techreport}. ESCS is standardized with an OECD mean of approximately 0 and a standard deviation of approximately 1. This variable is available for all 78 countries with minimal missingness (only Germany at 12.7\% and Mexico at 10.1\% exceed 10\%; no country exceeds 20\%). The weighted country-level ESCS means range from $-1.89$ (Morocco) to $+0.55$ (Iceland), reflecting the substantial socioeconomic diversity across participating countries. Students with missing ESCS values were excluded via listwise deletion, reducing the sample from 574,187 to 571,251 (0.5\% loss). The R and Stan code used for data preparation and estimation are available from the authors, and the Stan model code is reproduced in the online supplement (Section~S11).

\subsubsection{Achievement Gaps Across Countries}

A key motivation for applying SABDB to these data is the substantial heterogeneity in immigration-related achievement gaps across countries. Among the 32 countries with at least 200 students in both the native and first-generation subgroups, the mean first-generation achievement gap (native minus first-generation) is $+14.2$ points, but the standard deviation of this gap across countries is 48.0 points, with a range from $-95.7$ to $+91.1$ points. This enormous variation reflects fundamentally different immigration contexts. In Gulf states such as the United Arab Emirates and Qatar, immigrant students outperform natives by a large margin (gaps of $-95.7$ and $-81.6$ points, respectively), reflecting the high socioeconomic status of professional expatriate populations \parencite{OECD2018resilience}. In contrast, in countries such as Indonesia and B-S-J-Z China, native students outperform first-generation immigrants by over 100 points. Any model that assumes a single, fixed immigration effect across all 78 countries will necessarily misrepresent the majority of country-specific patterns.

\subsection{Model Specification}

For the PISA analysis, the SABDB design vector $\mathbf{x}_{sj}$ comprises an intercept, dummy variables for first- and second-generation immigration status (native as the reference category), and the ESCS index. Each country $s$ therefore has its own coefficient vector $\boldsymbol{\beta}_s = (\beta_{s,0}, \beta_{s,\text{1st}}, \beta_{s,\text{2nd}}, \beta_{s,\text{ESCS}})^\top$, giving it its own intercept, immigration effects, and SES gradient. The likelihood, the hierarchical prior with coefficient-specific between-country variances $\tau^2_{\beta,k}$, and the hyperpriors are exactly as specified for the SABDB model in Section~\ref{sec:sabdb_model} (Equations~\ref{eq:sabdb_likelihood}--\ref{eq:sigma_prior}). All variables are standardized to $z$-scores before estimation, and posterior summaries are transformed back to the PISA scale so that the $\tau_{\beta,k}$ are comparable across coefficients.

The HBSAE comparison model (Equation~\ref{eq:hbsae}) shares a single fixed coefficient vector $\boldsymbol{\beta}$ across all 78 countries and adds a country-specific random intercept $u_s \sim \mathcal{N}(0, \sigma_u^2)$, so only the intercept varies by country while the immigration and ESCS slopes are held identical everywhere. Both models are fit without survey weights under a model-based inference approach \parencite{gelman2007}, with the covariates serving as proxies for the factors driving differential selection, and both omit the school level, which would otherwise add roughly 21,000 parameters. Because that simplification is shared, the relative comparison is unaffected, and $\sigma_y^2$ absorbs both within- and between-school variation. The benchmark direct estimates, by contrast, are computed with survey weights via \texttt{EdSurvey} to ensure design-unbiased estimation.

Each PV-specific fit used Stan \parencite{carpenter2017stan} with 4 chains and adapt\_delta of 0.95. SABDB converged readily (maximum $\hat{R} \le 1.015$, minimum effective sample size $\ge 375$ across all 10 PV fits, zero divergent transitions, 3.6--7.9 hours per fit). HBSAE required extended 10,000-iteration runs for six PVs and still retained $\hat{R}$ between 1.05 and 1.09 for those fits, with runtimes of 11--56 hours (roughly 300 total hours vs.\ 50 for SABDB). The residual HBSAE convergence issues, if anything, favor HBSAE in the comparisons below, since incomplete mixing inflates posterior variance and widens intervals. Full computational and convergence tables appear in the supplement (Section~S7).

\subsection{Parameter Estimates and the Heterogeneity Structure}

Table~\ref{tab:pisa_params} presents global parameter estimates. The SABDB global means for the immigration effects are negative ($-5.6$ and $-3.2$ points), indicating that on average across the 78 countries immigrant students score slightly below native students after controlling for ESCS. The corresponding HBSAE coefficients, by contrast, are positive (6.2 and 0.7 points). This discrepancy arises because the HBSAE intercept absorbs country-level variation differently when slopes are fixed rather than country-specific. The SABDB between-country standard deviations reveal exactly the coefficient-specific heterogeneity structure the method is designed to exploit: country intercepts vary most ($\tau_\beta \approx 48$ PISA points), immigration effects vary substantially ($\tau_\beta \approx 29$--$39$ points), and the ESCS gradient is comparatively homogeneous ($\tau_\beta \approx 9$ points). SABDB therefore borrows strongly for the SES slope, which stabilizes small domains, while allowing immigration effects to vary across countries. HBSAE cannot make this distinction: between-country variation in slopes is forced into the residual ($\sigma_y$ rises from 82.7 to 83.7) and all domains are pulled toward a single global regression surface.

\begin{table}[ht]
\centering
\caption{Parameter Estimates: SABDB and HBSAE Models (PISA Scale)}
\label{tab:pisa_params}
\begin{threeparttable}
\begin{tabular}{lcccc}
\toprule
& \multicolumn{2}{c}{SABDB} & \multicolumn{2}{c}{HBSAE} \\
\cmidrule(lr){2-3} \cmidrule(lr){4-5}
Parameter & Estimate & SD & Estimate & SD \\
\midrule
\multicolumn{5}{l}{\textit{Fixed effects}} \\
\quad Intercept & 458.8 & 5.3 & 472.3 & 4.6 \\
\quad First-generation & $-5.6$ & 1.2 & 6.2 & 0.5 \\
\quad Second-generation & $-3.2$ & 0.9 & 0.7 & 0.5 \\
\quad ESCS & 35.1 & 1.0 & 31.5 & 0.1 \\
\midrule
\multicolumn{5}{l}{\textit{Variance components}} \\
\quad $\tau_{\beta,\text{Intercept}}$ / $\sigma_u$ & 47.6 & 4.0 & 45.9 & 3.8 \\
\quad $\tau_{\beta,\text{1st-Gen}}$ & 38.6 & 3.6 & --- & --- \\
\quad $\tau_{\beta,\text{2nd-Gen}}$ & 29.1 & 2.7 & --- & --- \\
\quad $\tau_{\beta,\text{ESCS}}$ & 8.9 & 0.7 & --- & --- \\
\quad $\sigma_y$ & 82.7 & 0.1 & 83.7 & 0.1 \\
\bottomrule
\end{tabular}
\begin{tablenotes}
\small
\item \textit{Note.} SABDB fixed effects are global means ($\mu_{\beta,k}$), whereas HBSAE fixed effects are pooled coefficients. SABDB variance components are between-country SDs per coefficient, whereas HBSAE has a single random-intercept SD ($\sigma_u$). All values in PISA points.
\end{tablenotes}
\end{threeparttable}
\end{table}

\subsection{Domain-Level Estimation: Coverage Analysis}

The primary evaluation of model performance compares the posterior credible intervals from SABDB and HBSAE against direct estimates computed from the survey data. Because the true population parameters are unknown in this empirical application (unlike the simulation study), we use direct estimates as the benchmark. Direct estimates are design-unbiased but can be highly imprecise for small domains; model-based estimates should improve precision while remaining consistent with the direct evidence.

Using direct estimates as the benchmark is a pragmatic choice, not an ideal one. Direct estimates are design-unbiased under the sampling design, but they are not the true population parameters. For small domains, direct estimates can be highly variable. So agreement between a model estimate and a noisy direct estimate does not guarantee that the model is close to the truth. However, a model estimate that disagrees with a noisy direct estimate might actually be more accurate. This is a limitation of any real-data evaluation where the true parameters are unknown, and it is why the simulation study presented earlier is still needed alongside this empirical analysis. The simulation lets us evaluate against known population values, while the PISA application tests the methods under realistic data complexity.

We evaluate model performance using the \textit{interval overlap rate}, which asks whether the model's 95\% credible interval and the direct estimate's 95\% confidence interval overlap \parencite{brown2001evaluation}. This criterion is appropriate when both the model estimate and the direct estimate carry uncertainty, as is the case for small domains where direct estimates are imprecise.

The direct 95\% confidence interval for each domain is $\bar{Q} \pm 1.96\sqrt{T}$, using the combined plausible-value estimate $\bar{Q}$ and total variance $T$ defined above. For the model-based estimates, the 95\% credible intervals are the 2.5th and 97.5th percentiles of the combined posterior draws across all 10 PV fits. Table~\ref{tab:pisa_coverage} presents the interval overlap results, stratified by domain sample size.

\begin{table}[ht]
\centering
\caption{Interval Overlap Rate: Model 95\% CI vs.\ Direct 95\% CI}
\label{tab:pisa_coverage}
\begin{threeparttable}
\begin{tabular}{lrcc}
\toprule
\textbf{Sample Size Group} & \textbf{Domains} & \textbf{SABDB} & \textbf{HBSAE} \\
\midrule
$N \leq 30$ & 32 & 96.9\% & 34.4\% \\
$30 < N \leq 60$ & 18 & 88.9\% & 22.2\% \\
$60 < N \leq 100$ & 16 & 100.0\% & 43.8\% \\
$100 < N \leq 200$ & 20 & 90.0\% & 30.0\% \\
$200 < N \leq 500$ & 27 & 85.2\% & 22.2\% \\
$N > 500$ & 121 & 33.1\% & 16.5\% \\
\midrule
All domains & 234 & 61.5\% & 23.1\% \\
\bottomrule
\end{tabular}
\begin{tablenotes}
\small
\item \textit{Note.} Each domain is a country-by-immigration-status combination (78 countries $\times$ 3 subgroups = 234 domains). All model estimates use combined 10-PV posteriors; direct estimates use Rubin's rules across 10 PVs.
\end{tablenotes}
\end{threeparttable}
\end{table}

\begin{figure}[H]
\centering
\includegraphics[width=0.85\textwidth]{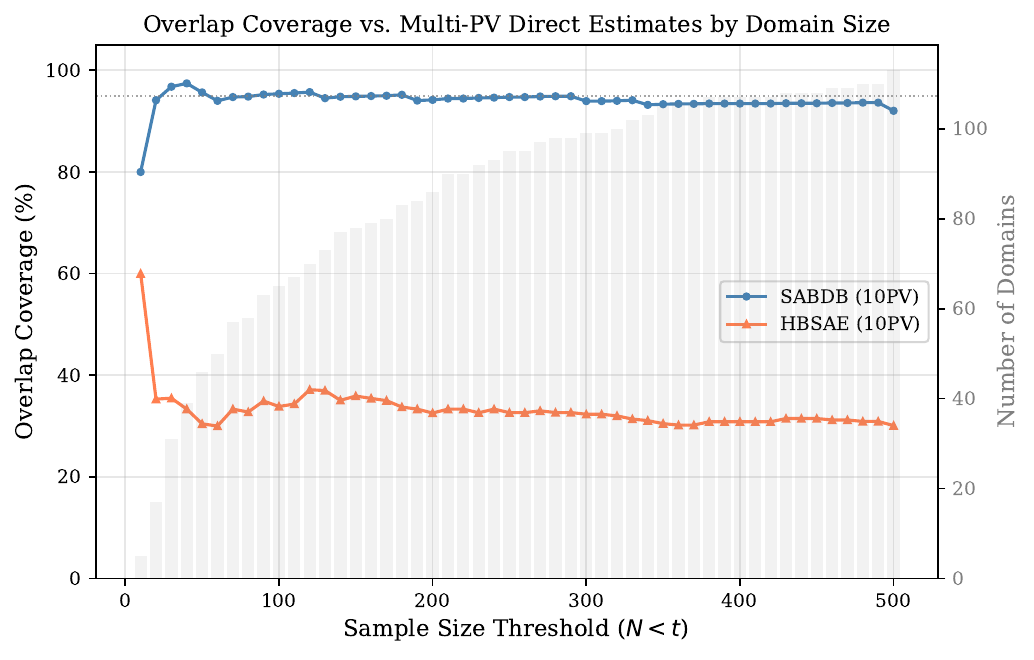}
\caption{Interval overlap rate as a function of domain sample size threshold. For each threshold $t$, the rate is computed over all domains with $N < t$. Gray bars indicate the number of domains included at each threshold.}
\label{fig:pisa_overlap}
\end{figure}

For the smallest domains ($N \leq 30$), SABDB achieves a 96.9\% interval overlap rate, meaning that in nearly all cases the model's credible interval is consistent with the direct survey evidence. HBSAE, by contrast, achieves only 34.4\% for these same domains. This pattern persists across all sample size strata. SABDB maintains overlap rates of 89--100\% for domains with fewer than 200 students, while HBSAE remains below 44\%.

The overall coverage figures (61.5\% for SABDB, 23.1\% for HBSAE) are lower because they include the large native-student domains where direct estimates are very precise and even small model-based deviations can fall outside the narrow direct confidence intervals. For the small domains that are the primary targets of small area estimation, SABDB's performance is excellent.. Domain-level plots for all 78 countries appear in the supplement (Section~S8).

\subsection{Credible Interval Width}

A complementary measure of model performance is the width of the credible intervals, which indicates how much uncertainty reduction the model achieves relative to direct estimation. Table~\ref{tab:pisa_ci_width} compares the average 95\% credible (or confidence) interval widths across the three estimation approaches.

\begin{table}[ht]
\centering
\caption{Average 95\% Interval Width (PISA Score Points) by Domain Sample Size}
\label{tab:pisa_ci_width}
\begin{tabular}{lrccc}
\toprule
\textbf{Sample Size Group} & \textbf{Domains} & \textbf{Direct} & \textbf{SABDB} & \textbf{HBSAE} \\
\midrule
$N \leq 30$ & 32 & 108.9 & 78.5 & 5.9 \\
$30 < N \leq 60$ & 18 & 71.5 & 55.5 & 6.3 \\
$60 < N \leq 100$ & 16 & 50.3 & 42.4 & 6.4 \\
$100 < N \leq 200$ & 20 & 39.7 & 32.4 & 6.2 \\
$200 < N \leq 500$ & 27 & 27.6 & 22.6 & 5.9 \\
$N > 500$ & 121 & 10.1 & 9.0 & 5.5 \\
\midrule
All domains & 234 & 35.6 & 27.9 & 5.8 \\
\bottomrule
\end{tabular}
\end{table}

Small area Bayesian dynamic borrowing substantially reduces interval width relative to direct estimation: for domains with $N \leq 30$, the average SABDB interval (78.5 points) is 28\% narrower than the direct interval (108.9 points), demonstrating meaningful efficiency gains from borrowing. At the same time, SABDB intervals appropriately widen for smaller domains, reflecting genuine uncertainty about domain means when sample sizes are limited. Narrower intervals can also reflect stronger model dependence rather than greater accuracy in any specific domain; the comparison documents the magnitude of borrowing-induced narrowing relative to direct estimation, not its correctness for any particular country-by-subgroup combination.

Hierarchical Bayesian small-area estimation intervals, by contrast, are nearly constant at approximately 5.8 points regardless of domain sample size. This uniformity is a diagnostic indicator of overconfidence. The model produces intervals that are far too narrow for small domains, failing to reflect the substantial uncertainty inherent in estimating a domain mean from as few as 2 or 4 observations. The HBSAE model's fixed-slope structure forces all domains toward the global regression surface, producing precise but potentially misleading estimates that do not account for the heterogeneity in immigration effects across countries.

\begin{figure}[H]
\centering
\includegraphics[width=0.85\textwidth]{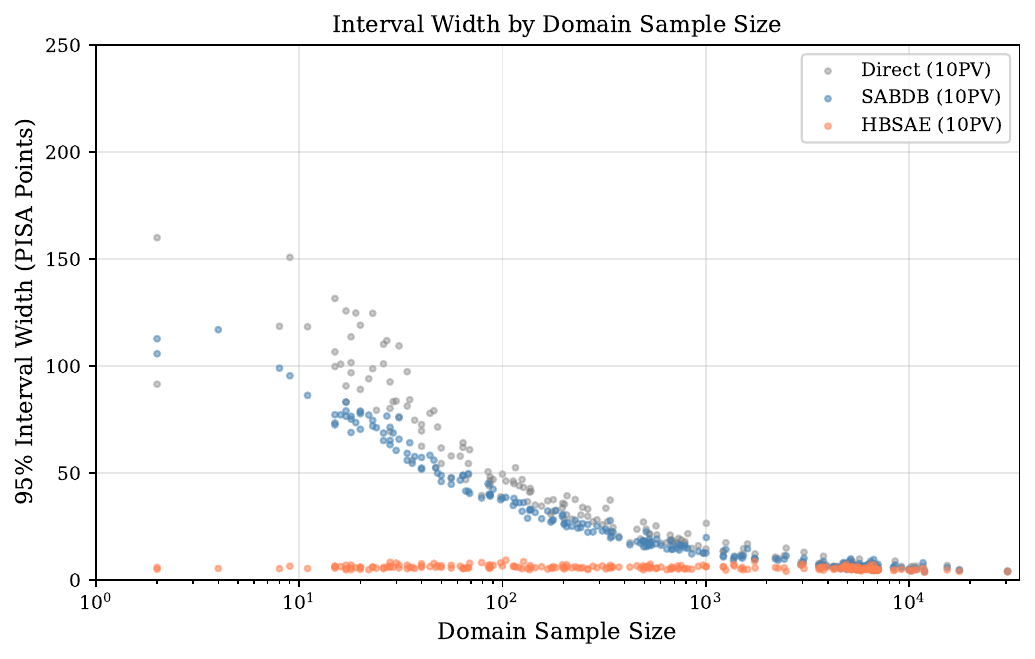}
\caption{95\% interval width by domain sample size for direct estimation, SABDB, and HBSAE. SABDB intervals widen appropriately for smaller domains while remaining narrower than direct intervals. HBSAE intervals are nearly constant regardless of sample size.}
\label{fig:pisa_ci_width}
\end{figure}

Figure~\ref{fig:pisa_ci_width} displays the 95\% interval width for each domain as a function of sample size, illustrating the contrasting behavior of the three estimation approaches. Direct intervals decrease with sample size as expected, SABDB intervals show a similar but attenuated pattern reflecting the efficiency gains from borrowing, and HBSAE intervals remain nearly flat regardless of domain size.

\subsection{Bias Relative to Direct Estimates}

For domains with sufficiently large sample sizes, direct estimates are precise enough to serve as a reliable benchmark against which to assess systematic bias in the model-based estimates. Table~\ref{tab:pisa_bias} presents the mean absolute difference (MAD) between each model's posterior mean and the direct estimate, stratified by domain sample size.

\begin{table}[H]
\centering
\caption{Mean Absolute Difference from Direct Estimates (PISA Points)}
\label{tab:pisa_bias}
\begin{threeparttable}
\begin{tabular}{lrccc}
\toprule
\textbf{Sample Size Group} & \textbf{Domains} & \textbf{SABDB} & \textbf{HBSAE} & \textbf{SABDB Closer (\%)} \\
\midrule
$N \leq 30$ & 32 & 32.8 & 78.5 & 90.6 \\
$30 < N \leq 60$ & 18 & 27.2 & 67.7 & 88.9 \\
$60 < N \leq 100$ & 16 & 14.3 & 33.0 & 87.5 \\
$100 < N \leq 200$ & 20 & 14.3 & 41.6 & 95.0 \\
$200 < N \leq 500$ & 27 & 15.1 & 35.6 & 85.2 \\
$N > 500$ & 121 & 14.9 & 21.4 & 67.8 \\
\midrule
All domains & 234 & 18.2 & 36.9 & 78.2 \\
\bottomrule
\end{tabular}
\begin{tablenotes}
\small
\item \textit{Note.} Mean absolute difference (MAD) is computed as $|$model estimate $-$ direct estimate$|$, averaged over domains in each group. ``SABDB Closer'' indicates the percentage of domains where SABDB's estimate is closer to the direct estimate than HBSAE's.
\end{tablenotes}
\end{threeparttable}
\end{table}

Across all 234 domains, SABDB produces estimates that are closer to the direct estimates in 78.2\% of cases, with a mean absolute difference of 18.2 points compared to 36.9 points for HBSAE. The advantage is most pronounced for domains with $30 < N \leq 60$ (MAD of 27.2 vs.\ 67.7) and for the smallest domains with $N \leq 30$ (MAD of 32.8 vs.\ 78.5), where HBSAE's fixed-slope constraint produces the largest deviations from the direct evidence. Even for the largest domains with $N > 500$, where direct estimates are precise, SABDB's MAD (14.9) is roughly 70\% of HBSAE's (21.4). This gap in HBSAE performance is expected, as the model constrains all 78 countries to share the same immigration and ESCS slopes, leaving no room to accommodate the cross-country variation in these effects.

Domain-level differences between each model's posterior mean and the direct estimate, plotted against sample size, appear in the online supplement (Section~S9). SABDB deviations cluster tightly around zero across the full range of sample sizes, whereas HBSAE shows large systematic deviations, particularly for small domains where the fixed-slope constraint forces estimates away from the direct evidence.

\subsection{Why HBSAE Underperforms: A Structural Explanation}\label{sec:pisa_why_hbsae}

The poor performance of HBSAE in this application is not a failure of Bayesian estimation or hierarchical modeling per se, but rather a consequence of structural misspecification. The standard HBSAE model assumes that the regression slopes for immigration status and ESCS are identical across all 78 countries, with only the intercept allowed to vary. This assumption is strongly violated in the PISA data. As documented in the preceding subsection on achievement gaps, the first-generation achievement gap ranges from $-96$ to $+91$ points across countries. By constraining this gap to a single global value, HBSAE effectively pulls all country-specific estimates toward the global regression surface. With 571,251 observations, the global estimates are extremely precise, so the model becomes overconfident in its predictions, producing narrow credible intervals that fail to capture the true domain means.

One might ask whether this problem could be resolved simply by adding random slopes to the HBSAE model, without invoking the dynamic borrowing framework. To address this question, we fitted an extended HBSAE model with country-specific coefficients for all four predictors, using standard half-Cauchy(0, 1) priors on the between-country standard deviations (i.e., the same structural flexibility as SABDB but without the Inverse-Gamma dynamic borrowing prior). This model produced results that were nearly identical to SABDB, with the interval overlap rate within 1--2 percentage points across all sample size strata. Approximate leave-one-out cross-validation supports the same conclusion and is reported in the online supplement (Section~S13).

This finding has important implications for interpreting the results. The improvement of SABDB over the standard HBSAE is attributable to the structural flexibility of allowing country-specific coefficients, not to the specific choice of prior on the between-country variance. With 78 countries, the data provide sufficient information to estimate the between-country variances precisely, and both the Inverse-Gamma and half-Cauchy priors yield essentially the same posterior. 

Given the similar performance between SABDB and a random slopes HBSAE model, one might then ask why use SABDB. The advantage of SABDB is that it provides a principled framework that naturally delivers this structural flexibility with built-in dynamic borrowing. Moreover, SABDB extends the Bayesian dynamic borrowing framework of \textcite{viele2014use} and \textcite{kaplan2023bayesian1, kaplan2023bayesian2} to the small area estimation context, where the between-source variance parameters govern the degree of information sharing and the prior on those parameters encodes the analyst's beliefs about commensurability. Simply adding random slopes to a standard hierarchical model indeed gives the same structural flexibility, but SABDB frames these parameters explicitly as borrowing parameters and the Inverse-Gamma prior is specifically chosen to regulate borrowing behavior. Furthermore, SABDB arrives at the country-specific coefficient structure by design, without requiring the analyst to first diagnose slope heterogeneity and then manually extend the model. The $\tau^2_{\beta,k}$ values provide direct posterior evidence about which coefficients require area-specific estimation and which can be safely pooled. In applications with fewer areas, where the data alone cannot precisely identify the between-area variances, the prior on $\tau^2_{\beta,k}$ has more influence on the posterior and the choice between the Inverse-Gamma dynamic borrowing prior and a standard weakly informative prior matters for both point estimates and interval coverage.

\subsection{Prior Sensitivity}

Robustness to the hyperprior on $\tau^2_\beta$ was assessed with five Inverse-Gamma specifications, IG(1, 0.001), IG(0.001, 0.001), IG(0.01, 0.01), IG(1, 0.1), and IG(1, 1), each fit as a full production run on the complete data (supplement, Table~S10). Four of the five yield virtually identical posteriors (intercept $\tau_\beta$ within 47.5--48.5 points; subgroup $\tau_\beta$ within 29.1--40.0). The exception, IG(1, 1), places substantial prior mass on large variances and roughly doubles the subgroup $\tau_\beta$ values, reducing borrowing and widening intervals. Even then the intercept $\tau_\beta$ moves only from 47.5 to 50.4 and $\sigma_y$ is unchanged at 82.66. With 78 areas the posterior is overwhelmingly data-driven for all but the most informative specification.

\section{Discussion}

Traditional SAE models treat borrowing as a fixed structural assumption, built into global variance components or exchangeability constraints. SABDB instead treats the degree of borrowing as a quantity governed by the posterior distribution of the between-area variance parameters $\tau^2_k$, learned jointly with all other model parameters through MCMC sampling. The $\tau^2_k$ parameters serve as coefficient-specific shrinkage regulators: when $\tau^2_k$ is small, the posterior for coefficient $k$ is pulled strongly toward the global mean (strong borrowing); when $\tau^2_k$ is large, each area retains its own estimate (weak borrowing). This mechanism allows SABDB to address long-standing concerns in the SAE literature about over-pooling and bias under heterogeneity \parencite{jiang2006mixed, rao2015small}.

A key contribution of this article is examining the exchangeability assumptions that underlie SAE models. Classical area-level models such as Fay-Herriot, and their Bayesian analogues, rely on a single global variance component to govern borrowing across all domains \parencite{fay1979estimates}. This structure implicitly assumes a level of homogeneity that is rarely justified in educational contexts, where subgroup effects can vary substantially across states or countries.

Small area Bayesian dynamic borrowing relaxes this assumption by allowing borrowing strength to vary across coefficients. In SABDB, borrowing is not hidden inside the model structure. Instead, it is an explicit part of the inference that can be measured and interpreted through the posterior distribution of $\tau^2_k$. The result is a form of partial exchangeability: countries are treated as exchangeable with respect to the ESCS gradient (where $\tau^2$ is small) but not with respect to immigration effects (where $\tau^2$ is large). This connects SABDB to the broader SAE literature on heterogeneity-aware modeling \parencite{rao2015small, jiang2006mixed} and to the dynamic borrowing framework developed in clinical trials \parencite{viele2014use} and extended to educational assessments by \textcite{kaplan2023bayesian1, kaplan2023bayesian2}.

The choice of NAEP as the motivating context is deliberate. NAEP is the nation's primary measure of educational achievement, and its data are frequently used to inform federal and state policy \parencite{NCES_NationsReportCard}. Yet the persistent non-reporting of subgroup outcomes limits NAEP's ability to fully serve this role. This article therefore aims both to advance statistical methodology and to support more equitable educational measurement.

This article developed SABDB, a unit-level small area estimation method that transfers the dynamic borrowing principle from historical data (borrowing across time) to small area domains (borrowing across areas), with coefficient-specific between-area variances governing the degree of pooling. Two evaluations, one against known truth under a conservative simulation design, one against design-based direct estimates in genuinely heterogeneous international data, yield a consistent picture.

First, SABDB's central property is robust interval calibration. In the simulation it maintained 0.93 coverage whether or not cross-area homogeneity held, while the standard HBSAE model's coverage depended entirely on an unverifiable structural assumption, collapsing to 0.50 (and to zero in particular domains) under heterogeneity. In PISA, SABDB intervals agreed with direct survey evidence for 97\% of the smallest domains while being 28\% narrower, whereas HBSAE produced 6-point intervals for domains with a handful of students. For reporting applications, where an interval that implies more certainty than the data warrant is arguably worse than no estimate at all, calibration is the property that matters most.

Second, the two studies jointly clarify when dynamic borrowing is beneficial. Under homogeneity with very small cells ($n \le 3$), heavy global shrinkage is more accurate on point estimates, and SABDB's adaptivity costs precision. No approach can compensate when the data carry almost no domain-level information. In domains that are small enough to be suppressed but not near-empty, and wherever genuine heterogeneity exists, SABDB matches or exceeds HBSAE on accuracy while preserving calibration. The PISA heterogeneity estimates ($\tau_\beta \approx 9$ points for the SES gradient vs.\ 29--48 for immigration effects and intercepts) illustrate the mechanism: the model discovered, rather than assumed, which aspects of the regression relationship transfer across countries.

Third, the PISA structural comparison marks the boundary of where SABDB's borrowing prior matters. With many areas, the data identify the between-area variances directly, so a conventional hierarchical model with random slopes and weakly informative priors recovers essentially SABDB's results. SABDB supplies that flexible structure by design and reports interpretable borrowing parameters by default. Its distinct advantage emerges in few-area regimes (such as borrowing among the 10--20 states relevant to a NAEP subgroup analysis), where the borrowing prior itself shapes inference, and where the dynamic borrowing tradition provides guidance \parencite{viele2014use, kaplan2023bayesian1}.

\subsection{Limitations and Future Directions}

Several limitations in this study should be noted.  Both studies fit unweighted models under a model-based inference framework \parencite{gelman2007}. This is defensible when the covariates in the model account for the factors driving differential selection, but incorporating survey weights directly into the Bayesian framework, through pseudo-likelihood or fully Bayesian weighted approaches, would strengthen the connection to design-based inference \parencite{you2003pseudo, si2020bayesian, parker2023comprehensive}. The PISA models omit the school level, so residual variance absorbs between-school variation. The simplification is shared by both models, but a three-level extension is feasible. The diagonal hierarchical prior ignores correlations among coefficients. A full covariance specification could improve borrowing further. The simulation's generative model used fixed slopes and random intercepts, which advantages HBSAE and means SABDB's simulation results are conservative, but also limits what the simulation reveals about SABDB under richer data-generating processes. In the empirical study, agreement with noisy direct estimates is an imperfect criterion: a model estimate that disagrees with a noisy direct estimate may be the more accurate of the two, which is why the simulation and empirical studies are needed jointly.

Several methodological extensions follow naturally. Traditional spatial small area models encode dependence through fixed geographic adjacency, yet in educational settings adjacent states may differ substantially while distant jurisdictions share similar subgroup patterns. Future models could replace fixed neighborhood structures with similarity learned jointly across covariates, space, and time, and could adapt temporal borrowing when structural breaks or policy changes introduce divergence across assessment cycles \parencite{bleiberg2023happened}. An area-level version of SABDB, for settings where only aggregate domain statistics are available, could estimate the borrowing structure directly within the posterior---through correlated random effects or adaptive shrinkage priors on area-specific variance components---rather than fixing it in advance. The framework also extends beyond subgroup means to proficiency rates, achievement gaps, growth measures, and composite indicators, and it invites probabilistic benchmarking strategies that preserve adaptive borrowing while ensuring coherence across levels of aggregation.

On the inferential and computational side, embedding survey-weighted likelihoods directly within the framework would bridge the design-based and model-based inference traditions \parencite{you2003pseudo, parker2023comprehensive}. Because the amount of borrowing is governed by the between-area variance parameters $\tau^2_k$, whose prior can strongly influence results when the number of areas is small, future work could develop sensible default priors for these parameters together with diagnostics that let applied researchers verify how much borrowing is taking place. Finally, approximate Bayesian inference methods such as variational inference or integrated nested Laplace approximation could scale SABDB to national assessment production environments involving large numbers of domains and time points.

Despite these limitations, the combined evidence supports SABDB as a robust default for unit-level small area estimation in educational assessments: it matches the calibration of conventional models when their assumptions hold, remains well calibrated when those assumptions fail, reveals the heterogeneity structure it relies on, and is computationally cheaper than the conventional alternative at scale.

\section*{Acknowledgments}
This article is based on the first author's doctoral dissertation at the University of Wisconsin--Madison. The first author thanks the dissertation committee for their guidance. Computation was performed on the Center for High Throughput Computing (CHTC) at the University of Wisconsin--Madison. This manuscript is under review at the \emph{Journal of Educational and Behavioral Statistics}.

\singlespacing
\printbibliography[title={References}]

@article{bleiberg2023happened,
  title={What happened to the K--12 education labor market during COVID? The acute need for better data systems},
  author={Bleiberg, Joshua F and Kraft, Matthew A},
  journal={Education Finance and Policy},
  volume={18},
  number={1},
  pages={156--172},
  year={2023},
  publisher={MIT Press One Rogers Street, Cambridge, MA 02142-1209, USA journals-info~…}
}

@book{snijders2012multilevel,
  title     = {Multilevel Analysis: An Introduction to Basic and Advanced Multilevel Modeling},
  author    = {Snijders, Tom A. B. and Bosker, Roel J.},
  edition   = {2},
  year      = {2012},
  publisher = {SAGE}
}

@unpublished{yavuz2024absenteeismncme,
  title   = {Potential Effect of Absenteeism on Declines of Students’ Performance in NAEP 2022},
  author  = {Yavuz, Sinan and Kim, Y. Y.},
  note    = {Presented at the National Council on Measurement in Education (NCME), April 11--14, 2024, Philadelphia, PA},
  year    = {2024}
}

@misc{NCES_ScoresAchv,
  author    = "{National Center for Education Statistics}",
  title     = "{NAEP Scores and Achievement Levels}",
  year      = {2025},
  url       = {https://nces.ed.gov/nationsreportcard/guides/scores_achv.aspx},
  urldate   = {2025-09-17}
}

@misc{NCES_MathConstants_2022,
  author    = "{National Center for Education Statistics}",
  title     = "{Transformation Constants Used for the 2022 NAEP Mathematics Assessment}",
  year      = {2022},
  url       = {https://nces.ed.gov/nationsreportcard/tdw/analysis/2022/trans_constants_math2022.aspx},
  urldate   = {2025-09-17}
}

@misc{NCES_NDE,
  author    = "{National Center for Education Statistics}",
  title     = "{NAEP Data Explorer}",
  publisher = "{U.S. Department of Education, Institute of Education Sciences}",
  year      = {2025},
  url       = {https://www.nationsreportcard.gov/ndecore/},
  urldate   = {2025-09-16}
}

@misc{NCES_School_Locations,
  author    = "{National Center for Education Statistics}",
  title     = "{School Locations}",
  year      = {2024},
  url       = {https://nces.ed.gov/programs/edge/Geographic/SchoolLocations},
  urldate   = {2024-09-12}
}

@misc{NCES_CCD_Data_Files,
  author    = "{National Center for Education Statistics}",
  title     = "{Common Core of Data (CCD) - CCD Data Files}",
  year      = {2024},
  url       = {https://nces.ed.gov/ccd/files.asp},
  urldate   = {2024-09-12}
}

@article{reardon2024separate,
  title={Is separate still unequal? New evidence on school segregation and racial academic achievement gaps},
  author={Reardon, Sean F and Weathers, Ericka S and Fahle, Erin M and Jang, Heewon and Kalogrides, Demetra},
  journal={American Sociological Review},
  volume={89},
  number={6},
  pages={971--1010},
  year={2024},
  publisher={SAGE Publications Sage CA: Los Angeles, CA}
}

@techreport{krenzke2020piaac,
title={Program for the International Assessment of Adult Competencies ({PIAAC}): State and County Estimation Methodology Report. NCES 2020-225},
author={Krenzke, Tom and Mohadjer, Leyla and Li, Jianzhu and Erciulescu, Andreea and Fay, Robert and Ren, Weijia and Van de Kerckhove, Wendy and Li, Lin and Rao, J. N. K.},
institution={National Center for Education Statistics},
year={2020}
}

@article{yang2024relationship,
  title={The relationship between mathematics self-efficacy and mathematics achievement: multilevel analysis with {NAEP} 2019},
  author={Yang, Yao and Maeda, Yukiko and Gentry, Marcia},
  journal={Large-Scale Assessments in Education},
  volume={12},
  number={1},
  pages={16},
  year={2024},
  publisher={Springer}
}

@article{parker2023comprehensive,
  title={A comprehensive overview of unit-level modeling of survey data for small area estimation under informative sampling},
  author={Parker, Paul A and Janicki, Ryan and Holan, Scott H},
  journal={Journal of Survey Statistics and Methodology},
  volume={11},
  number={4},
  pages={829--857},
  year={2023},
  publisher={Oxford University Press}
}

@misc{NCES_NationsReportCard,
  author = {{National Center for Education Statistics}},
  title = {{The Nation's Report Card}},
  year = {2025},
  url = {https://nces.ed.gov/nationsreportcard/},
  note = {Accessed: 2025-03-04}
}

@article{efron1977stein,
  title={Stein's paradox in statistics},
  author={Efron, Bradley and Morris, Carl},
  journal={Scientific American},
  volume={236},
  number={5},
  pages={119--127},
  year={1977},
  publisher={JSTOR}
}

@article{hobbs2012commensurate,
  title={Commensurate priors for incorporating historical information in clinical trials using general and generalized linear models},
  author={Hobbs, Brian P and Sargent, Daniel J and Carlin, Bradley P},
  journal={Bayesian Analysis (Online)},
  volume={7},
  number={3},
  pages={639},
  year={2012}
}

@article{hobbs2011hierarchical,
  title={Hierarchical commensurate and power prior models for adaptive incorporation of historical information in clinical trials},
  author={Hobbs, Brian P and Carlin, Bradley P and Mandrekar, Sumithra J and Sargent, Daniel J},
  journal={Biometrics},
  volume={67},
  number={3},
  pages={1047--1056},
  year={2011},
  publisher={Oxford University Press}
}

@article{ibrahim2000power,
  title={Power prior distributions for regression models},
  author={Ibrahim, Joseph G and Chen, Ming-Hui},
  journal={Statistical Science},
  pages={46--60},
  year={2000},
  publisher={JSTOR}
}

@article{pocock1976combination,
  title={The combination of randomized and historical controls in clinical trials},
  author={Pocock, Stuart J},
  journal={Journal of chronic diseases},
  volume={29},
  number={3},
  pages={175--188},
  year={1976},
  publisher={Elsevier}
}

@article{edwards2024using,
  title={Using {b}ayesian dynamic borrowing to maximize the use of existing data: a case-study},
  author={Edwards, Dawn and Best, N and Crawford, J and Zi, L and Shelton, C and Fowler, A},
  journal={Therapeutic Innovation \& Regulatory Science},
  volume={58},
  number={1},
  pages={1--10},
  year={2024},
  publisher={Springer}
}

@misc{NCES2025,
  author = {{National Center for Education Statistics}},
  title = {Summary Rules for Minimum Sample Sizes for Reporting Results},
  year = {2025},
  url = {https://nces.ed.gov/nationsreportcard/tdw/analysis/summary_rules_minimum.aspx},
  urldate = {2025-02-28},
  organization = {U.S. Department of Education}
}

@book{jones2004nation,
  title={The nation's report card: evolution and perspectives},
  author={Jones, Lyle V and Olkin, Ingram},
  year={2004},
  publisher={Phi Delta Kappa Educational Foundation}
}

@article{viele2014use,
  title={Use of historical control data for assessing treatment effects in clinical trials},
  author={Viele, Kert and Berry, Scott and Neuenschwander, Beat and Amzal, Billy and Chen, Fang and Enas, Nathan and Hobbs, Brian and Ibrahim, Joseph G and Kinnersley, Nelson and Lindborg, Stacy and others},
  journal={Pharmaceutical statistics},
  volume={13},
  number={1},
  pages={41--54},
  year={2014},
  publisher={Wiley Online Library}
}

@article{arora1997superiority,
  title={On the superiority of the Bayesian method over the BLUP in small area estimation problems},
  author={Arora, Vipin and Lahiri, Partha},
  journal={Statistica Sinica},
  pages={1053--1063},
  year={1997},
  publisher={JSTOR}
}

@article{fay1979estimates,
  title={Estimates of income for small places: an application of James-Stein procedures to census data},
  author={Fay III, Robert E and Herriot, Roger A},
  journal={Journal of the American Statistical Association},
  volume={74},
  number={366a},
  pages={269--277},
  year={1979},
  publisher={Taylor \& Francis}
}

@book{rao2015small,
  title={Small area estimation},
  author={Rao, John NK and Molina, Isabel},
  year={2015},
  publisher={John Wiley \& Sons}
}

@article{gelman2007,
author = {Andrew Gelman},
title = {Struggles with Survey Weighting and Regression Modeling},
volume = {22},
journal = {Statistical Science},
number = {2},
publisher = {Institute of Mathematical Statistics},
pages = {153 -- 164},
year = {2007},
doi = {10.1214/088342306000000691},
URL = {https://doi.org/10.1214/088342306000000691}
}

@article{si2020bayesian,
  title={Bayesian hierarchical weighting adjustment and survey inference},
  author={Si, Yajuan and Trangucci, Rob and Gabry, Jonah Sol and Gelman, Andrew},
  journal={Survey Methodology},
  pages={181},
  year={2020}
}

@article{you2003pseudo,
  title={Pseudo hierarchical {B}ayes small area estimation combining unit level models and survey weights},
  author={You, Yong and Rao, JNK},
  journal={Journal of Statistical Planning and Inference},
  volume={111},
  number={1-2},
  pages={197--208},
  year={2003},
  publisher={Elsevier}
}

@inproceedings{ghosh1992hierarchical,
  title={A hierarchical {B}ayes approach to small area estimation with auxiliary information},
  author={Ghosh, Malay and Lahiri, Parthasarathi},
  booktitle={Bayesian Analysis in Statistics and Econometrics},
  pages={107--125},
  year={1992},
  organization={Springer}
}

@article{jiang2006mixed,
  title={Mixed model prediction and small area estimation},
  author={Jiang, Jiming and Lahiri, Partha},
  journal={Test},
  volume={15},
  pages={1--96},
  year={2006},
  publisher={Springer}
}

@article{kaplan2023bayesian1,
  title={Bayesian dynamic borrowing of historical information with applications to the analysis of large-scale assessments},
  author={Kaplan, David and Chen, Jianshen and Yavuz, Sinan and Lyu, Weicong},
  journal={Psychometrika},
  volume={88},
  number={1},
  pages={1--30},
  year={2023},
  publisher={Springer}
}

@article{kaplan2023bayesian2,
  title={Bayesian historical borrowing with longitudinal large-scale assessments},
  author={Kaplan, David and Chen, Jianshen and Lyu, Weicong and Yavuz, Sinan},
  journal={Large-scale Assessments in Education},
  volume={11},
  number={1},
  pages={1--30},
  year={2023},
  publisher={SpringerOpen}
}

@article{mislevy1992estimating,
  title={Estimating population characteristics from sparse matrix samples of item responses},
  author={Mislevy, Robert J and Beaton, Albert E and Kaplan, Bruce and Sheehan, Kathleen M},
  journal={Journal of Educational Measurement},
  volume={29},
  number={2},
  pages={133--161},
  year={1992},
  publisher={Wiley Online Library}
}

@misc{rubin1987multiple,
  title={Multiple imputation for survey nonresponse},
  author={Rubin, Donald B},
  year={1987},
  publisher={New York: Wiley}
}

@book{PISA2018,
   author = {OECD},
   title = {{PISA 2018} {R}esults: ({V}olumes {I-IV}): What Students Know and Can Do},
   year = {2019},
   publisher = {OECD Publishing},
   address = {Paris},
   url = {https://doi.org/10.1787/5f07c754-en}
}

@manual{edsurvey,
  title = {{EdSurvey}: Analysis of {NCES} Education Survey and Assessment Data},
  author = {Paul Bailey and Ahmad Emad and Huade Huo and Michael Lee and Yuqi Liao and Alex Lishinski and Trang Nguyen and Qingshu Xie and Jiao Yu and Ting Zhang and Eric Buehler and Jeppe Bundsgaard and Ren C'deBaca and Sinan Yavuz},
  year = {2024},
  note = {R package version 4.0.7},
  url = {https://CRAN.R-project.org/package=EdSurvey}
}

@article{battese1988error,
  title={An error-components model for prediction of county crop areas using survey and satellite data},
  author={Battese, George E and Harter, Rachel M and Fuller, Wayne A},
  journal={Journal of the American Statistical Association},
  volume={83},
  number={401},
  pages={28--36},
  year={1988},
  publisher={Taylor \& Francis}
}

@article{ZhouReiter,
  title={A note on {B}ayesian inference after multiple imputation},
  author={Zhou, Xiang and Reiter, Jerome P},
  journal={The American Statistician},
  volume={64},
  number={2},
  pages={159--163},
  year={2010},
  publisher={Taylor \& Francis}
}

@book{pisa2018techreport,
  author = {OECD},
  title = {{PISA 2018 Technical Report}},
  publisher = {OECD Publishing},
  address = {Paris},
  year = {2019},
  url = {https://www.oecd.org/pisa/data/pisa2018technicalreport/}
}

@book{OECD2018resilience,
  author = {OECD},
  title = {The Resilience of Students with an Immigrant Background: Factors that Shape Well-being},
  publisher = {OECD Publishing},
  address = {Paris},
  year = {2018},
  doi = {10.1787/9789264292093-en}
}

@book{PISA2018vol2,
  author = {OECD},
  title = {{PISA 2018 Results (Volume II): Where All Students Can Succeed}},
  publisher = {OECD Publishing},
  address = {Paris},
  year = {2019},
  doi = {10.1787/b5fd1b8f-en}
}

@article{wu2005plausible,
  author = {Wu, Margaret},
  title = {The role of plausible values in large-scale surveys},
  journal = {Studies in Educational Evaluation},
  volume = {31},
  number = {2--3},
  pages = {114--128},
  year = {2005}
}

@article{carpenter2017stan,
  author = {Carpenter, Bob and Gelman, Andrew and Hoffman, Matthew D. and Lee, Daniel and Goodrich, Ben and Betancourt, Michael and Brubaker, Marcus and Guo, Jiqiang and Li, Peter and Riddell, Allen},
  title = {Stan: A probabilistic programming language},
  journal = {Journal of Statistical Software},
  volume = {76},
  number = {1},
  pages = {1--32},
  year = {2017}
}

@inproceedings{brown2001evaluation,
  author    = {Brown, G. and Chambers, R. and Heady, P. and Heasman, D.},
  title     = {Evaluation of Small Area Estimation Methods: An Application to Unemployment Estimates from the {UK LFS}},
  booktitle = {Proceedings of Statistics Canada Symposium 2001: Achieving Data Quality in a Statistical Agency: A Methodological Perspective},
  year      = {2001},
  publisher = {Statistics Canada}
}

@book{PISA2022vol2,
  author = {OECD},
  title = {{PISA 2022 Results (Volume II): Learning During -- and From -- Disruption}},
  year = {2023},
  publisher = {OECD Publishing},
  address = {Paris},
  doi = {10.1787/a97db61c-en}
}

@article{smith1995bayesian,
  title={Bayesian approaches to random-effects meta-analysis: A comparative study},
  author={Smith, Thomas C and Spiegelhalter, David J and Thomas, Andrew},
  journal={Statistics in Medicine},
  volume={14},
  number={24},
  pages={2685--2699},
  year={1995},
  publisher={Wiley}
}

\end{document}